\documentclass[english]{article}
\pdfoutput=1
\usepackage[T1]{fontenc}
\usepackage[latin9]{inputenc}
\usepackage{babel}
\usepackage{authblk}
\usepackage{graphicx}
\usepackage{fullpage}

\author[1]{Evan L. Runnerstrom\thanks{Equal contribution}}
\author[1]{Kyle P. Kelley\thanks{Equal contribution}}
\author[2]{Thomas G. Folland}
\author[3]{Nader Engheta}
\author[2]{Joshua D. Caldwell\thanks{josh.caldwell@vanderbilt.edu}}
\author[1,4]{Jon-Paul Maria\thanks{jpm133@psu.edu}}
\affil[1]{Department of Materials Science and Engineering,
North Carolina State University, Raleigh, NC 27695}
\affil[2]{Department of Mechanical Engineering, Vanderbilt
University, Nashville, TN 37212}
\affil[3]{Department of Electrical and Systems Engineering,
University of Pennsylvania, Philadelphia, PA 19104}
\affil[4]{Department of Materials Science and Engineering,
The Pennsylvania State University, University Park, PA 16802}
\date{}

\title{Polaritonic hybrid-epsilon-near-zero modes: engineering strong optoelectronic
coupling and dispersion in doped cadmium oxide bilayers }

\begin{document}
\maketitle
\begin{abstract}
Polaritonic materials that support epsilon-near-zero (ENZ) modes offer the opportunity to design light-matter interactions at the nanoscale through phenomena like resonant perfect absorption and extreme sub-wavelength light concentration. To date, the utility of ENZ modes is limited in propagating polaritonic systems by a relatively flat spectral dispersion, which gives ENZ modes small group velocities and short propagation lengths. Here we overcome this constraint by coupling ENZ modes to surface plasmon polariton (SPP) modes in doped cadmium oxide ENZ-on-SPP bilayers. What results is a strongly coupled hybrid mode, characterized by strong anti-crossing and a large spectral splitting on the order of 1/3 of the mode frequency. The resonant frequencies, dispersion, and coupling of these polaritonic-hybrid-epsilon-near-zero (PH-ENZ) modes are controlled by tailoring the modal oscillator strength and the ENZ-SPP spectral overlap, which can potentially be utilized for actively tunable strong coupling at the nanoscale. PH-ENZ modes ultimately leverage the most desirable characteristics of both ENZ and SPP modes through simultaneous strong interior field confinement and mode propagation.
\end{abstract}
\section{Introduction}
Extraordinary light-matter interactions become possible when the effective dielectric permittivity of a material vanishes. Such epsilon-near-zero (ENZ) phenomena occur near the frequency where the real part of the dielectric function changes sign and are characterized by electromagnetic modes with light wavelengths in the ENZ medium diverging toward infinity, along with decoupled spatial and temporal electromagnetic fields.\cite{Liberal:2017uh} Harnessing this behavior has recently become a major objective in nanophotonics research; examples include resonant perfect light absorption in deeply sub-wavelength thin films,\cite{Luk:2014kw, Campione:2015bs, Campione:2015kc, Runnerstrom:2017ee} length-invariant antenna resonances,\cite{Kim:2016kx} wave-front engineering,\cite{Alu:2007bm} controlled thermal emissivity,\cite{Dewalt:2013ji, Liberal:2018ew} and extraordinary transmission by supercoupling.\cite{Silveirinha:2006ed} Practically, an ENZ condition can be realized in any material or system where the real part of the effective dielectric function passes through zero, and thus occurs naturally in materials such as doped semiconductors\cite{Campione:2015bs, Runnerstrom:2017ee, Traviss:2013fg} and polar crystals.\cite{Vassant:2012dr, Vassant:2012iz, Caldwell:2015fp, Nordin:2017cd} Thin layers of these materials can additionally support polaritonic ``ENZ modes'', which strongly confine light and electric fields within the interior of deeply sub-wavelength films, leading to resonant perfect absorption.\cite{Luk:2014kw, Campione:2015bs, Campione:2015kc} ENZ modes have the additional distinguishing feature of being dispersion-less, as their resonant energy does not change with the longitudinal component of the incident wavevector. Consequently, ENZ modes have a very low group velocity and are not ideal when energy propagation is desired. Propagating ENZ behavior can be engineered within artificially-structured metamaterials,\cite{Maas:2013es} though these metamaterials typically use metallic components with high optical losses and require complicated and laborious fabrication procedures.\cite{Maas:2013es} As such, there is clear motivation to combine low-loss natural ENZ materials within layered or otherwise engineered structures, which offers the potential to control the modal dispersion, propagation lengths, and operating frequencies in nanophotonic devices that exploit ENZ modes.

Doped semiconductors can naturally sustain spectrally tunable ENZ behavior near the screened plasma frequency, especially in the infrared (IR).\cite{Runnerstrom:2017ee, Traviss:2013fg,Taylor:2012fd} However, when most materials are doped to resonate at energies of interest in the near- to mid-IR, they suffer a steep drop in carrier mobility due to ionized impurity scattering, which prohibitively increases optical losses. Doped cadmium oxide (CdO) is an important exception, as thin films are shown to support low-loss plasmonic modes over a broad spectral range pursuant to tunable carrier concentrations that range from 10\textsuperscript{19} to 10\textsuperscript{21} e\textsuperscript{-}/cm\textsuperscript{3} while maintaining  carrier mobilities over 300 cm\textsuperscript{2}/V·s (peak values >500 cm\textsuperscript{2}/V·s).\cite{Runnerstrom:2017ee, Sachet:2015ho, Kelley:2017ee} Optically, CdO supports tunable, low-loss surface plasmon polaritons (SPPs) and ENZ modes over the entire mid-IR.\cite{Runnerstrom:2017ee, Sachet:2015ho} However, as discussed above, low group velocities limit the implementation of ENZ films in applications such as waveguide design, where some of the extraordinary behaviors would be most beneficial. Conversely, SPP modes propagate at much faster group velocities, but the strong interfacial confinement renders them highly sensitive to surface morphology and carrier scattering losses.\cite{Bozhevolnyi:2002eg} Here, we demonstrate a strategy to simultaneously overcome the shortcomings of SPP and ENZ modes by coupling SPP and ENZ resonances into a new class of hybrid modes within a monolithic bilayer system.

Coupling between photonic and/or electronic states\cite{Vahala:2003cx, Campione:2016gz, Jun:2013hh} occurs when the two constituent states interact with sufficient strength to form a hybrid system where energy is exchanged between the states more quickly than it decays out of the system. Such a coupling interaction between two degenerate energy states changes the spatial and spectral dispersion of the two modes and causes the modes to repel, resulting in symmetric and antisymmetric modes with a well-defined splitting in energy (analogous to the vacuum Rabi splitting).\cite{Torma:2015is} Notably, this changes the properties of both modes and introduces new physics associated with the hybrid mode. Among the most well-known examples are coupling between high quality optical resonators and excitons in dye molecules, J-aggregates, and quantum dots.\cite{Torma:2015is, Hakala:2009iz, Schlather:2013gc, Gomez:2010iz} Similar effects have also been observed in split ring resonators coupled to plasmonic modes and vibrational states coupled to optical cavities.\cite{Campione:2016gz, Jun:2013hh, Dunkelberger:2016gi, Simpkins:2015kr, Shalabney:2015gm} Here, we demonstrate analogous strong coupling between SPP and ENZ modes in tunable, homoepitaxial CdO ENZ-on-SPP bilayers. The coupling is characterized by a very large and tunable spectral splitting that is on the order of the mode frequency itself. The result is a hybrid of both SPP and ENZ modes---with high field confinement characteristic of the ENZ and fast mode propagation characteristic of the SPP---which we call a polaritonic-hybrid-epsilon-near-zero or ``PH-ENZ'' mode. We demonstrate that coupling in PH-ENZ modes is highly tunable by varying either the carrier concentration and/or thickness of the individual layers. This work identifies and explores new ways to combine ENZ and SPP behaviors in propagating nanophotonic systems built around the concept of monolithic metamaterial structures.
\section{Results and Discussion}
In the following, we investigate a series of ENZ-on-SPP bilayers to explore ENZ-SPP coupling. All of the bilayer structures studied here commonly contain a 300 nm-thick indium-doped CdO (In:CdO) heteroepitaxial layer grown on r-plane sapphire using reactive high-power impulse magnetron sputtering (R-HiPIMS). This In:CdO layer, hereafter referred to as the ``SPP layer'', has an electron concentration of $\sim$2.5$\times$10\textsuperscript{20}~cm\textsuperscript{-3}, a mobility of $\sim$375 cm\textsuperscript{2}/V·s, and exhibits strong wavevector-dependent absorption due to polaritonic dispersion across the mid-IR when probed using the Kretschmann configuration. These reflectance measurements exhibit the typical asymptotic dispersion $\left[\frac{k}{k_0}=n_\mathrm{prism}\sin\left(\theta\right)\right]$ characteristic of an SPP mode (Fig. 1a). When the film thickness is reduced well below the optical skin depth of In:CdO, the SPP modes at the top and bottom film interfaces hybridize into the well-known long- and short-range (symmetric/high energy and anti-symmetric/low energy) SPP modes. In the extremely thin limit (about $50\times$ smaller than the plasma wavelength),\cite{Campione:2015kc} the dispersion of the symmetric mode is asymptotically pinned along the light line and the screened plasma frequency, resulting in a nominally flat spectral dispersion near the ENZ condition and confinement of the electric field to the film interior: this is the ENZ mode.\cite{Campione:2015kc, Runnerstrom:2017ee} We observe ENZ mode behavior in the dispersion relation for a 100 nm-thick In:CdO layer (the ``ENZ layer'', Fig. 1b) doped to  $\sim$8.1$\times$10\textsuperscript{19}~e\textsuperscript{-}/cm\textsuperscript{3} ($\mu\sim 400$ cm\textsuperscript{2}/V·s). Comparing the spectral dispersions of SPP and ENZ modes side by side in Fig. 1a-b leads to a natural question: given their very strong spectral overlap, how will these SPP and ENZ modes interact when placed in direct proximity? We explore this question by fabricating a bilayer structure through homoepitaxial growth of an ENZ layer on top of an SPP layer (identical thicknesses and carrier densities as in the isolated cases, see insets of Fig. 1a-c), which dramatically modifies the spectral dispersion (Fig. 1c).  We emphasize that bilayer growth requires no lithography, patterning, or other micro-/nanofabrication (see Fig. S1 for structural characterization of ENZ-on-SPP bilayers).

\begin{figure}
  \centering
  \includegraphics[width=\linewidth]{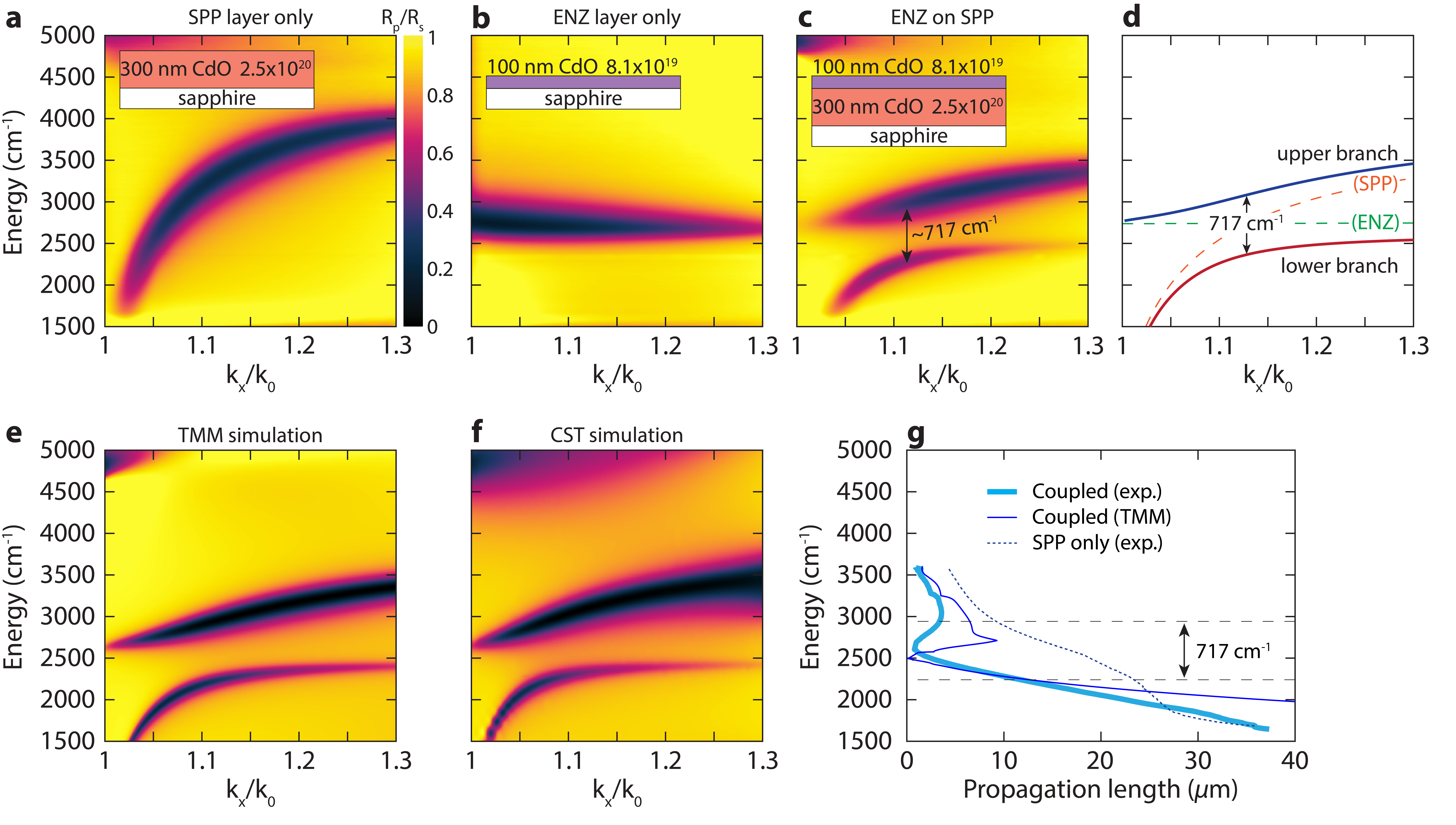}
  \caption{a) IR reflectivity spectral map (Kretschmann configuration) of a bare SPP mode. Inset: geometry/electron concentration of the SPP layer. b) Reflectivity map of a bare ENZ mode.  Inset: geometry/electron concentration of the ENZ layer. c) Reflectivity map of a coupled PH-ENZ mode, with the coupling point and splitting energy noted.  Inset: geometry/electron concentration of the bilayer. d) Theoretical dispersion curves of coupled upper/lower branches (thick blue/red lines) resulting from strong coupling between bare SPP/ENZ dispersion curves (dashed orange/green lines). e) Simulated (transfer matrix method) reflectivity map of the coupled ENZ-on-SPP bilayer. f) Simulated (CST) reflectivity map of the coupled ENZ-on-SPP bilayer. g) Calculated propagation lengths for the experimental SPP dispersion in part (a), the experimental coupled modes in part (c), and the simulated coupled modes in part (e). Note: reflectivity maps plot $R= R_p/R_s$ (the ratio of reflected p-polarized to s-polarized light) on the color axis, energy (wavenumbers) on the y-axis, and the film-parallel component of incident wavevector (normalized by $k_0=\omega/c$ at each frequency) on the x-axis. See Fig. S2 for a comparison between angle, wavevector, and normalized wavevector for the dispersion relation plotted in part (c).}
  \label{fig:1}
\end{figure}

\renewcommand{\thefootnote}{\fnsymbol{footnote}}
The bilayer sample exhibits clear anti-crossing between the ENZ and SPP modes (Fig. 1c), indicated by the splitting of the dispersion into upper and lower branches that acquire dispersive character over a significant portion of the measured spectral range. The anti-crossing occurs at the spectral location where the `bare' ENZ and SPP modes are tuned to cross (see Fig. 1d), which opens up a spectral transparency window. This is a clear signature of strong coupling between the two electromagnetic modes that produces a 717 cm\textsuperscript{-1} splitting (Fig. 1c and 1d). To confirm this interpretation, we model strong coupling between an ENZ and SPP mode using the following expression:\cite{Torma:2015is, Schlather:2013gc}

\begin{equation}
E_{\mathrm{coupled}}^{\mathrm{UB,LB}}(k_x)=\frac{E_{\mathrm{SPP}}(k_x)+E_{\mathrm{ENZ}}(k_x)}{2}\pm\frac{1}{2}\left[\Omega_{\mathrm{R}}^2+\left(E_{\mathrm{SPP}}(k_x)-E_{\mathrm{ENZ}}(k_x)\right)^2\right]^{\frac{1}{2}}
\end{equation} where $E_{\mathrm{coupled}}^{\mathrm{UB,LB}}(k_x)$ is the dispersion relation of the coupled upper and lower branches (UB/LB), $E_{\mathrm{SPP}}(k_x)$ is the dispersion relation of the uncoupled SPP mode\footnote{$k_x=\frac{\omega}{c}\left[\frac{\varepsilon_1\varepsilon_2}{\varepsilon_1+\varepsilon_2}\right]^{1/2}$, where $\varepsilon_1$ is is the effective dielectric constant of the dielectric medium (here, set to 2 to roughly account for the combined presence of the ENZ layer and air above the SPP layer), and $\varepsilon_2(\omega)=\varepsilon_\infty-\frac{\omega_p^2}{\omega^2+i\gamma\omega}$ is the dielectric function of the plasmonic layer ($\varepsilon_\infty$, $\omega_p$, $\gamma$: high frequency dielectric constant, plasma frequency, damping frequency).}, $E_{\mathrm{ENZ}}(k_x)$ is the dispersion relation of the uncoupled ENZ mode\footnote{$\omega\approx\omega_p\left[1-\frac{k_xd}{4}\right]-i\frac{\gamma}{2}$, where $d$ is film thickness, after Ref. \cite{Campione:2015kc}.}, and $\Omega_{\mathrm{R}}$ is the splitting energy\footnote{The spatial overlap between the ENZ and SPP electric fields: $\Omega(\mathbf{r})\propto\int\mathbf{E}_{\mathrm{ENZ}}(\mathbf{r})\cdot\mathbf{E}_{\mathrm{SPP}}(\mathbf{r})dV$,(Ref. \cite{Schlather:2013gc}), here simply set to 717 cm\textsuperscript{-1}}. This model is presented in Fig. 1d and quantitatively matches our experimental results well, supporting our interpretation that a hybrid PH-ENZ mode evolves through polaritonic strong coupling. Qualitatively, this behavior can be described as follows: distinct upper and lower branches (Fig. 1d blue and red curves, respectively) result from this coupling, with spectral splitting at the intersection of the bare SPP and bare ENZ dispersion curves (orange and green dashed curves, plotted for visual reference). At the coupling point, the UB (LB) is associated with a hybrid symmetric (antisymmetric) PH-ENZ mode that has both SPP and ENZ character. Spectrally removed from the coupling point, the UB and LB asymptotically approach the bare SPP and bare ENZ dispersion curves, respectively, and the mode profiles primarily adopt the behavior of the asymptote to which the branch is closest. In addition to the clear anti-crossing we observe, strong coupling requires that the following inequality is satisfied:\cite{Torma:2015is, Schlather:2013gc}

\begin{equation}
\Omega_{\mathrm{R}}>\frac{1}{2}\left(\gamma_{\mathrm{SPP}}+\gamma_{\mathrm{ENZ}}\right)
\end{equation} where $\gamma_{\mathrm{SPP}}$  and $\gamma_{\mathrm{ENZ}}$  are the damping frequencies of the SPP and ENZ layers, respectively. This implies that the two modes share resonant energy at a rate faster than their average decay rate. Thus, it appears that using high mobility In:CdO as the low-loss building block for our bilayers is an important prerequisite for the strong coupling we observe here. Note that the splitting energy is on the order (here, greater than 1/4) of the resonant energy at the coupling point, which may even place this effect within the ultrastrong coupling limit (where strong coupling is defined as spectral splitting greater than the mode linewidth, and ultrastrong coupling is defined as spectral splitting having the same order of magnitude as the resonant energy).\cite{Torma:2015is}

We are also able to accurately simulate the hybrid PH-ENZ mode of this same ENZ-on-SPP bilayer using two different computational methods, a MATLAB-based transfer matrix method (TMM, Fig. 1e) and a commercial finite-element technique (CST Studio, Fig. 1f). Agreement between our simulations and experimental results demonstrates the utility of both TMM and CST simulations to interpret the physical attributes of PH-ENZ modes. For completeness, we calculate the frequency-dependent imaginary part of the wavevector of the coupled system (Fig. S3), which provides insight into the loss characteristics and propagation lengths of coupled PH-ENZ modes. This analysis reveals that PH-ENZ modes have non-negligible propagation lengths on the order of 5-10 $\mu$m, (Fig. 1g) which is about 30-60\% of the SPP propagation length at the same frequencies. This `lossy' propagation is suitable for integrating ENZ behavior within waveguide-based nanophotonics, which would be difficult to achieve using uncoupled ENZ modes. 

To further understand PH-ENZ mode tunability, we grew a number of ENZ-on-SPP bilayers with systematically decreasing carrier concentration in the 100 nm-thick ENZ layer, effectively tuning or detuning the SPP-ENZ interaction and spectral overlap. Our experimental (top row) and calculated (bottom row) spectra are provided in Fig. 2 a-d and demonstrate that changing the ENZ frequency of the top layer directly controls the spectral overlap and coupling between the ENZ and SPP dispersions. As expected, when the ENZ layer has a much higher electron concentration than the SPP layer, there is no spectral overlap between the ENZ and SPP dispersion relations and little to no coupling results (Fig. 2a). As the carrier density decreases, the ENZ condition spectrally approaches the asymptote of the SPP dispersion, which occurs in this system once the carrier concentration of the ENZ layer drops to $\sim$1.5$\times$10\textsuperscript{20} e\textsuperscript{-}/cm\textsuperscript{3} (Fig. 2b).  At this point, ENZ-SPP hybridization and coupling occurs. As the electron concentration continues to decrease, the spectral overlap and the coupling point shift to lower energy and wavevector (Fig. 2c,d). Additionally, the upper branch acquires greater dispersive character suggesting that the UB and LB are strongly coupled over a wider range of wavevectors. We note that while tuning in this experiment is achieved by controlling the CdO ENZ layer carrier concentration through doping, dynamic tuning of PH-ENZ modes could, in principle, be realized through electrostatic gating or optical pumping approaches, as was recently demonstrated for localized surface phonon polariton nanostructures\cite{Dunkelberger:2018kz} and for In:CdO ENZ layers.\cite{Yang:2017hf}

\begin{figure}
  \centering
  \includegraphics[width=\linewidth]{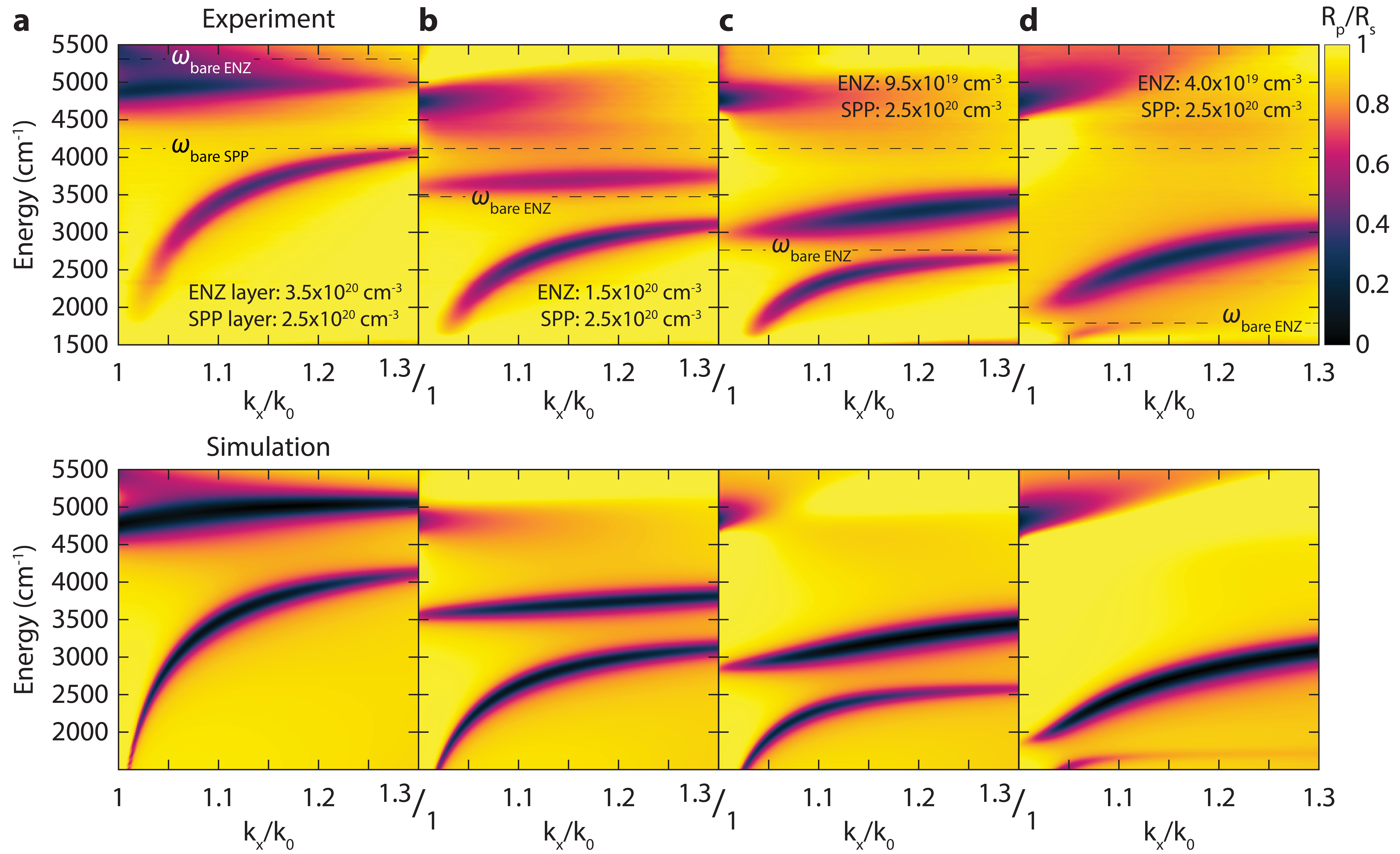}
  \caption{Tuning PH-ENZ modes by changing electron concentration in the ENZ layer. Top row: experimental IR reflectivity maps. Bottom row: TMM simulations. a) ENZ layer electron concentration: 3.5$\times$10\textsuperscript{20} e\textsuperscript{-}/cm\textsuperscript{3}. b) ENZ layer electron concentration: 1.5$\times$10\textsuperscript{20} e\textsuperscript{-}/cm\textsuperscript{3}. c) ENZ layer electron concentration: 9.5$\times$10\textsuperscript{19} e\textsuperscript{-}/cm\textsuperscript{3}. d) ENZ layer electron concentration: 4.0$\times$10\textsuperscript{19} e\textsuperscript{-}/cm\textsuperscript{3}. The ENZ layer is 100 nm thick. The SPP layer is 300 nm thick and has an electron concentration of 2.5$\times$10\textsuperscript{20} e\textsuperscript{-}/cm\textsuperscript{3}. The bare SPP and ENZ frequencies are denoted as dashed lines for reference.}
  \label{fig:2}
\end{figure}

A second critical element controlling the coupling strength is the oscillator strength of the constituent modes. This can be readily tuned by changing the ENZ layer thickness at a constant carrier density: as the thickness changes, so will the magnitude of the electric field overlap integral that determines the spectral splitting in the hybrid dispersion. For this experiment, we grew two additional ENZ-on-SPP bilayers, comparable to that in Fig. 2b, but one with an 80 nm-thick ENZ layer and the second with a 20 nm-thick ENZ layer (both ENZ layers have a carrier density of 1.5$\times$10\textsuperscript{20} e\textsuperscript{-}/cm\textsuperscript{3}). Experimental dispersion relationships (Fig. 3 a and d) of these bilayers show a clear and systematic decrease in the spectral splitting between the UB and LB as the ENZ layer thickness is reduced. Once again, both TMM (Fig. 3b,e) and CST simulations (Fig. 3c,f) capture this behavior.

\begin{figure}
  \centering
  \includegraphics[width=\linewidth]{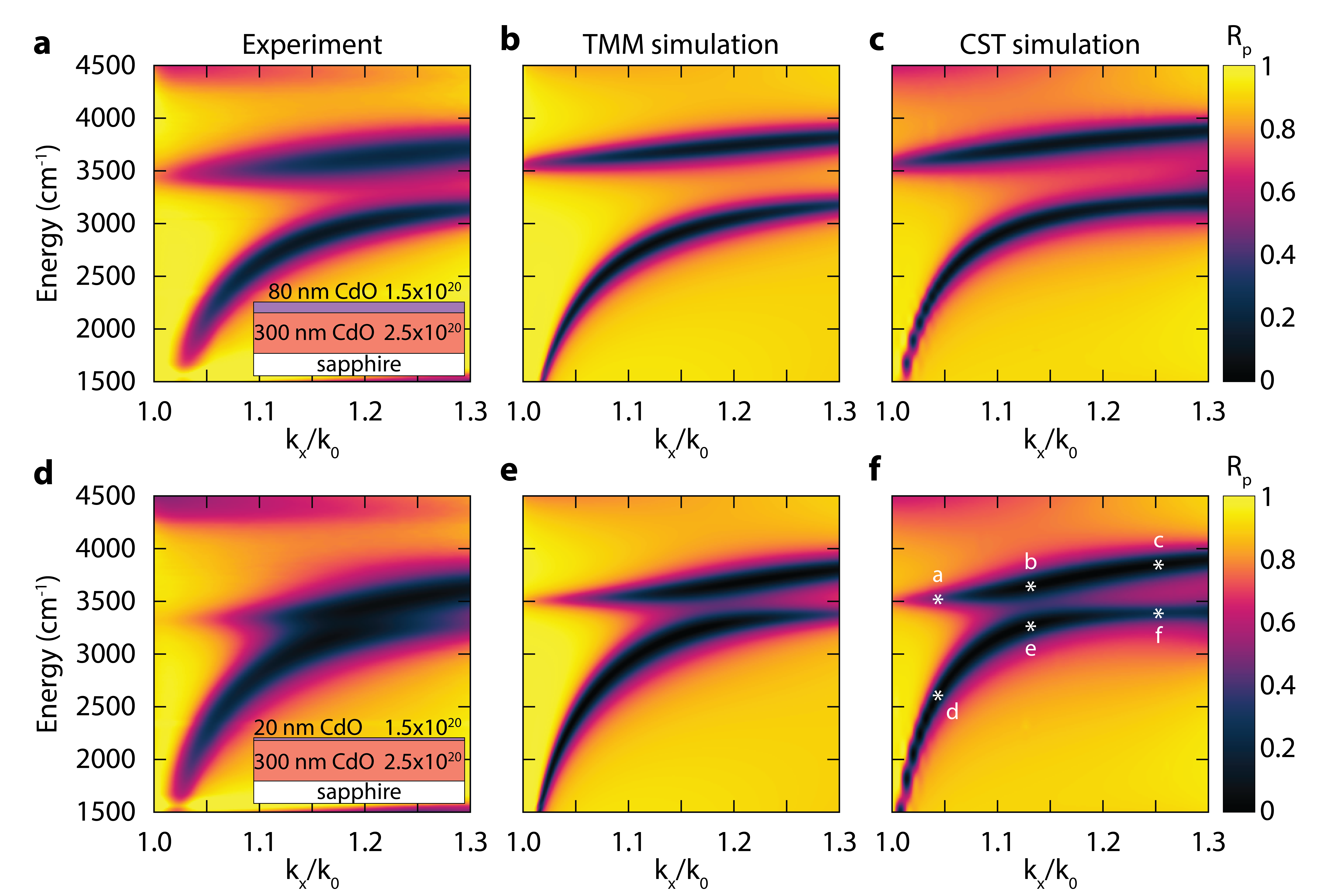}
  \caption{Tuning PH-ENZ modes in ENZ-on-SPP bilayers by changing ENZ layer thickness.  a) Experimental IR reflectivity map of an ENZ-on-SPP bilayer with an 80 nm thick ENZ layer. Inset: geometry/electron concentration of the bilayer. b) Simulated (TMM) reflectivity map of the structure in the inset of (a). c) Simulated (CST) reflectivity map of the structure in the inset of (a). d) Experimental IR reflectivity map of an ENZ-on-SPP bilayer with a 20 nm thick ENZ layer. Inset: geometry/electron concentration of the bilayer. e) Simulated (TMM) reflectivity map of the structure in the inset of (d). f) Simulated (CST) reflectivity map of the structure in the inset of (d). The asterisks denote the points simulated in Figure 4.}
  \label{fig:3}
\end{figure}

To investigate how hybridization influences the near-field behavior of PH-ENZ modes, we use CST simulations to visualize the optical modes and electric field profiles of the 20 nm ENZ / 300 nm SPP bilayer (Fig. 3d inset) at discrete points along the dispersion branches labeled a-f in Fig. 3f. These field profiles (electric field magnitude in the transverse/film normal/z-direction) and Poynting vectors are provided in Fig. 4a-f, where each specific panel corresponds to the position in Fig. 3f of the same letter label. The corresponding longitudinal/x-direction electric field profiles are provided in Fig. S4a-f. We analyze this particular structure because the regions of the dispersion relationship where both the UB and LB exhibit evidence of strong coupling (Fig. 4b,e) are clearly delineated from the regions where the modes are uncoupled (Fig. 4a,d) and/or weakly coupled (Fig. 4c,f).

\begin{figure}
  \centering
  \includegraphics[width=\linewidth]{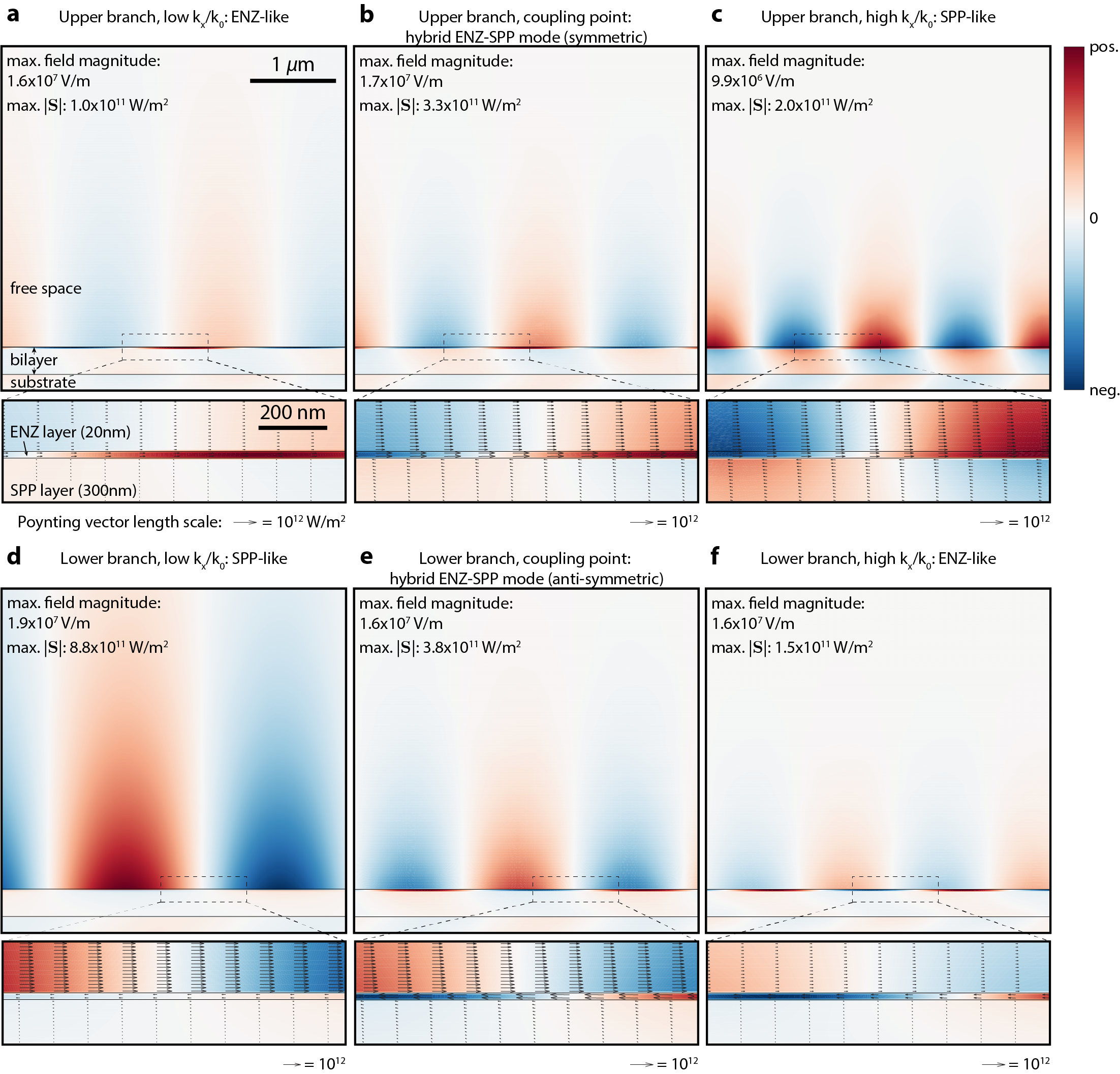}
  \caption{Electric field/mode profiles of the ENZ-on-SPP bilayer simulated in Figure 3f (see structure in Figure 3d inset). The maximum field amplitude and Poynting vector amplitude is denoted for each profile. a) Uncoupled upper branch mode profile (point a in Fig. 3f) with predominant ENZ character. Detail view: Field confinement within the 20 nm ENZ layer with low propagation. Poynting vector field is overlaid in all detail views. b) Strongly coupled upper branch PH-ENZ mode profile (point b in Fig. 3f) with hybrid ENZ-SPP behavior. The field profile is symmetric/in phase. Detail view: Symmetric simultaneous field confinement within the ENZ layer and evanescent decay into the free space above with propagation in the ENZ layer and at the bilayer surface. c) Uncoupled upper branch mode profile (point c in Fig. 3f) with predominant SPP character. Detail view: Evanescent field decay into free space and into the bulk of the SPP layer. d) Uncoupled lower branch mode profile (point d in Fig. 3f) with predominant SPP character. Detail view: Evanescent field decay into free space, with the propagating mode confined to the bilayer surface. e) Strongly coupled lower branch PH-ENZ mode profile (point e in Fig. 3f) with hybrid ENZ-SPP behavior. This field profile is antisymmetric/out of phase. Detail view: Antisymmetric simultaneous field confinement within the ENZ layer and evanescent decay into the free space above, with ENZ layer propagation in the opposite direction as at the bilayer surface. f) Uncoupled lower branch mode profile (point f in Fig. 3f) with predominant ENZ character. Detail view: Field confinement and low propagation within the 20 nm ENZ layer. Note that, while the mode profiles in the pairs of panels a/d, b/e, and c/f are simulated for the same $k_x/k_0$, the actual $k_x$ for each panel is unique due to the normalization factor $k_0$, which is unique at each frequency. Hence, the mode profile periodicity (proportional to non-normalized $k_x$) is slightly different in each panel.}
  \label{fig:4}
\end{figure}

At low wavevectors, the SPP and ENZ modes have sufficient spectral separation to limit significant coupling. Thus, we anticipate that these positions along the UB (Fig. 4a) and LB (Fig. 4d) should largely resemble the uncoupled ENZ and SPP modes, respectively. Indeed, the electric fields in Fig. 4a are largely confined within the 20 nm thick ENZ layer, as in an uncoupled ENZ mode. Likewise, the power flow (Poynting vector) is largely confined within the ENZ layer and the magnitude of the energy flux is low. In contrast, the LB at this wavevector (Fig. 4d) exhibits an optical mode that is confined at the bilayer surface with electric fields that evanescently decay into the surrounding medium, resembling an uncoupled SPP mode. For this SPP-like mode, the Poynting vector is directed along the bilayer surface and in the surrounding medium with correspondingly long propagation.

As the two branches converge and reach the anti-crossing point, they couple strongly.  At this point (Fig. 4b,e for the UB and LB, respectively), we observe modal profiles that simultaneously resemble both ENZ and SPP modes. In the UB (Fig. 4b), the symmetric PH-ENZ mode exhibits strong field confinement within the ENZ layer with simultaneous evanescent decay into the surrounding medium. Power flow is significantly modified, with the energy flux distributed throughout the ENZ layer and surrounding medium, and, to a small extent, the SPP layer. Importantly, power flow is greatest within the ENZ layer itself and is more than three times greater in this PH-ENZ mode than in the ENZ-only mode (compare Fig. 4a and 4b detail views). We calculate propagation lengths on the order of 5 $\mu$m (Fig. 1g) for this coupled PH-ENZ mode in the upper branch, thanks to the fact that the electric fields are no longer entirely confined to the ENZ layer and spectral dispersion is induced. The lower branch (Fig. 4e) consists of the anti-symmetric PH-ENZ mode, which again exhibits simultaneously strong evanescent fields and field confinement within the ENZ layer, but with the fields in the two layers out of phase. Here, the energy flux is again mostly confined within the ENZ layer. Thanks to the larger maximum Poynting vector magnitude, propagation distances are higher for this anti-symmetric PH-ENZ mode (Fig. 1g).

Finally, at wavevectors well above the strong coupling point (Fig. 4c,f), the UB, which initially began with ENZ character, transitions to SPP-like behavior (Fig. 4c), while the opposite is observed in the LB (original SPP character transitions to ENZ behavior with limited power flow and slow propagation; Fig. 4f). Interestingly, the UB at high wavevectors exhibits electric field decay into the CdO SPP layer and substrate (Fig. 4c); because of the low carrier densities (relative to metals) employed here, the skin depth at these frequencies is on the order of hundreds of nanometers to microns,\cite{Runnerstrom:2017ee} so the SPP-like mode can decay into both the dielectric ambient as well as into the CdO film itself.

We conclude our investigation with additional TMM simulations to thoroughly characterize how the tuning parameters influence coupling and the strength of spectral splitting in ENZ-on-SPP bilayers. To do so, we simulate the reflectivity spectra for bilayers with varying ENZ layer carrier concentration and thickness and determine the splitting by finding the smallest UB/LB peak separation. Fig. 5a displays the results of this analysis, with the splitting plotted against the detuning energy, which is defined here as the difference in plasma frequency between the SPP and ENZ layers, for a number of ENZ layer thicknesses. Two trends are immediately apparent. First, the splitting is maximized at detuning energies ranging from 2000-4000 cm\textsuperscript{-1}, which corresponds to ENZ layer carrier concentrations of 1--1.5$\times$10\textsuperscript{20} e\textsuperscript{-}/cm\textsuperscript{3}. As seen above, this detuning energy is consistent with the strongest overlap of the bare ENZ and bare SPP dispersion curves (SPP layer: 2.5$\times$10\textsuperscript{20} e\textsuperscript{-}/cm\textsuperscript{3}, 300 nm). The slight shift of the splitting maximum to higher detuning energies with ENZ layer thickness is likely the result of the increasing thickness changing the effective dielectric constant `felt' by the SPP layer (i.e., an effective medium), which suppresses the bare SPP dispersion curve and requires a slightly lower ENZ layer carrier concentration for maximum overlap. The second clear trend is that the splitting monotonically and categorically increases with ENZ layer thickness/oscillator strength as predicted (Fig. 5b). The fact that the ENZ layer carrier concentration strongly modulates the spectral splitting, especially in very thin (5-50 nm) ENZ layers, further indicates that future dynamic control of the ENZ layer's electronic properties will enable actively tunable PH-ENZ modes.

\begin{figure}
  \centering
  \includegraphics{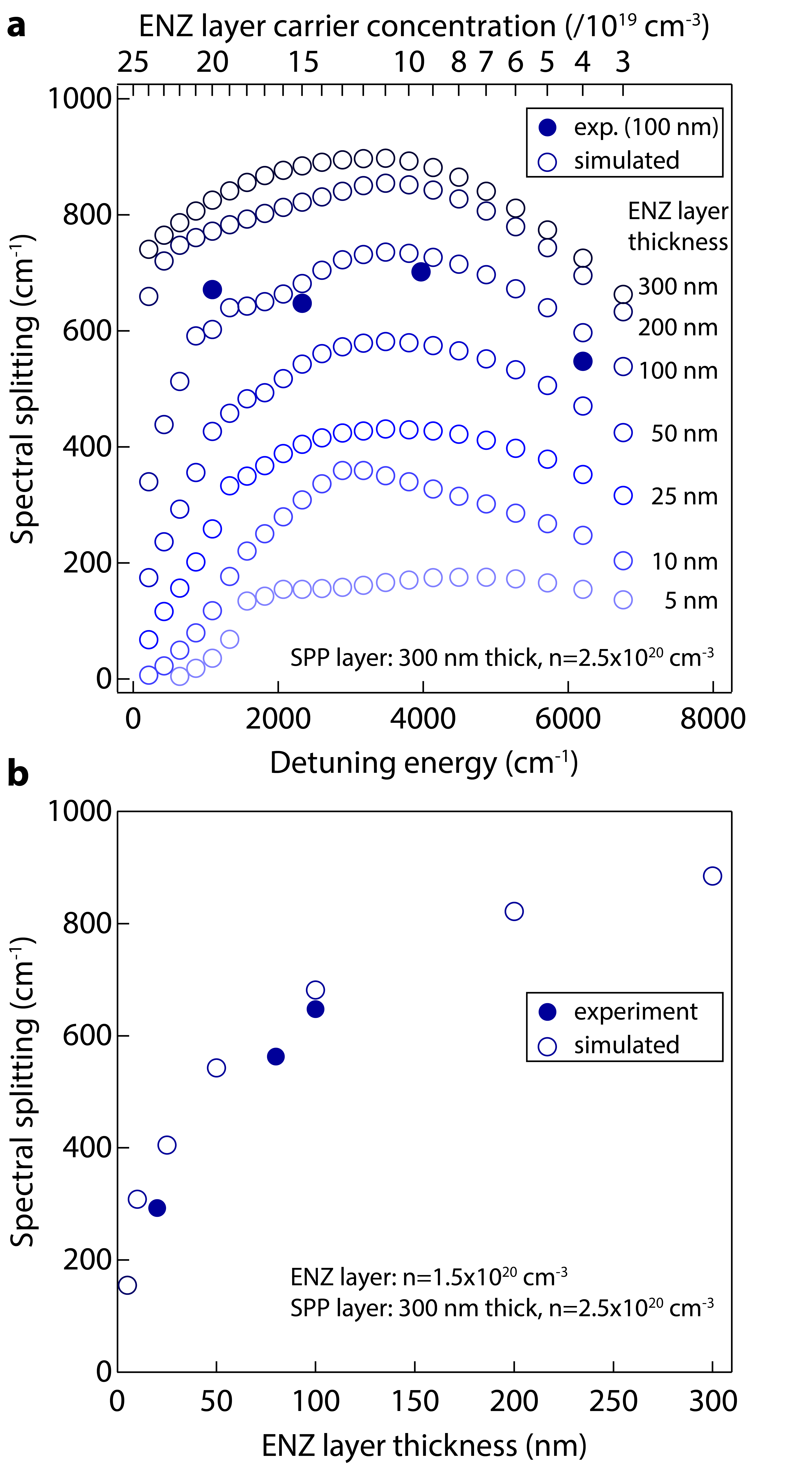}
  \caption{TMM simulations of coupling and spectral splitting for a wide range of ENZ layer electron concentration and ENZ layer thickness. a) PH-ENZ mode spectral splitting vs. detuning energy, which is defined as the difference between the SPP layer's plasma frequency and the ENZ layer's plasma frequency. Experimental results (filled circles) for ENZ-on-SPP bilayers with 100 nm thick ENZ layers are included for comparison. b) PH-ENZ mode spectral splitting vs. ENZ layer thickness. The ENZ carrier concentration is set at 1.5$\times$10\textsuperscript{20} e\textsuperscript{-}/cm\textsuperscript{3}. Experimental results (filled circles) are included for comparison.}
  \label{fig:5}
\end{figure}

\section{Conclusion}
We have demonstrated that ENZ-on-SPP interactions within bilayer In:CdO structures exhibit characteristics of (ultra)strong coupling as indicated by strong spectral splitting of the constituent dispersion curves: the experimentally-observed splitting can exceed 1/4 to 1/3 of the incident field energy. The resulting hybrid SPP-ENZ modes, or PH-ENZ modes, combine the most attractive features of SPP and ENZ modes by simultaneously achieving increased group velocity and electric field confinement within the bilayer. We believe that tunable, coupled PH-ENZ modes will be particularly attractive for applications in IR nanophotonic circuitry and waveguides, where simultaneous field confinement and mode propagation around uneven shapes and sharp corners is desirable. Because PH-ENZ modes are highly tailorable through changes to the ENZ layer thickness (electric field overlap) and the carrier concentration of either the ENZ or SPP layer (spectral overlap between the two modes) they could, in principle, be made dynamic in actively tunable configurations. This suggests that additional potential applications for strongly coupled PH-ENZ modes include quantum information processing or thresholdless lasing,\cite{Torma:2015is, Chikkaraddy:2016ja, Tame:2013iv, Cao:2018ce} and that the resonant absorption properties of these structures may also be highly useful for narrow-band thermal emitters.\cite{Greffet:2002jx, Wang:2017bb} Finally, the tunable nature of PH-ENZ modes offers the opportunity to tailor modal energies and dispersion to resonate with multiple specific electronic levels or vibrational states for improved surface-enhanced infrared sensing and catalysis.\cite{Cao:2018ce, Hutchison:2012dk, Agrawal:2017de, Yoo:2018dx, Autore:2018dn} Combined with the ease with which our bilayers are fabricated, we expect PH-ENZ modes to quickly find applications in one or more of these areas.

\section{Experimental}
Extensive descriptions of our methods for depositing doped CdO thin films can be found in the Supporting Information. Briefly, we deposit heteroepitaxial CdO onto epi-ready double side polished r-plane sapphire using reactive high-power impulse magnetron sputtering (R-HiPIMS) from a 99.9999\% pure metallic Cd target. Doping is achieved by RF co-sputtering from a 99.99\% pure metallic In target. Thickness/deposition rate and carrier concentration are calibrated using x-ray reflectivity and Hall effect measurements, respectively.  To form doped CdO bilayers, we first deposit 300 nm In:CdO (the SPP layer) with an electron concentration of 2.5$\times$10\textsuperscript{20} e\textsuperscript{-}/cm\textsuperscript{3}. After a 2 minute rest without breaking vacuum, a homoepitaxial In:CdO layer (the ENZ layer) is deposited on top of the SPP layer. The electron concentration and thickness of the ENZ layer are controlled by the RF power applied to the In target and the deposition time, respectively. Post-deposition, the bilayers are annealed at 700°C in pure oxygen for 30 minutes. X-ray diffraction confirms that the ENZ-on-SPP bilayers are homo-/heteroepitaxial, and secondary ion mass spectroscopy confirms that the In\textsuperscript{3+} donors are localized to the SPP and ENZ layers in different concentrations, as expected.

Optical properties (polarized reflectivity spectra) were characterized using a Woollam IR-VASE ellipsometer. A 90° calcium fluoride (CaF\textsubscript{2}) prism and $n$=1.720 index matching fluid were used to couple IR light into our plasmonic films through the polished substrate in the Kretschmann-Raether configuration. All reflectivity data were recorded with p- and s-polarized light and are plotted using $R=R_p/R_s$ as the color/z-coordinate, energy (in cm\textsuperscript{-1}) as the y-coordinate, and normalized wavevector as the x-coordinate ($k_x/k_0$, unitless, where $k_x$ is the component of the incident light wavevector parallel to the film surface, and $k_0=\omega/c$).

To simulate the optical response of our bilayer films, we use a combination of the transfer matrix model and the CST Studio software. Our TMM model is based on a homebuilt Matlab code that includes frequency-dependent dielectric dispersions for CaF\textsubscript{2} and sapphire and generates frequency-dependent dielectric functions for each plasmonic layer based on the Drude model. This provides a computationally-inexpensive method to simulate the reflectivity spectra of our coupled bilayers. The CST software solves Maxwell's equations using the finite element method, allowing us to simulate electric field profiles in addition to reflectivity spectra.

\section{Supporting Information}
Detailed descriptions of thin film growth, XRD and ToF-SIMS structural/compositional characterization of bilayers, additional dispersion relations in angle and un-normalized wavevector space, description and additional plot of complex wavevector calculations and propagation lengths, descriptions of TMM and CST simulations, additional electric field profiles, Poynting vector profiles.

\section{Acknowledgements}
We gratefully acknowledge support for this work by NSF grant CHE-1507947, by Army Research Office grants W911NF- 16-1-0406 and W911NF-16-1-0037, and by Office of Naval Research grant N00014-18-2107. TF and JDC both acknowledge support from Vanderbilt School of Engineering through the latter's startup funding package. NE acknowledges the partial support from the Vannevar Bush Faculty Fellowship program sponsored by the Basic Research Office of the Assistant Secretary of Defense for Research and Engineering and funded by the Office of Naval Research through grant N00014-16-1-2029.  In addition, this work was performed in part at the Analytical Instrumentation Facility (AIF), which is supported by the State of North Carolina and the National Science Foundation (award number ECCS-1542015). The AIF is a member of the North Carolina Research Triangle Nanotechnology Network (RTNN), a site in the National Nanotechnology Coordinated Infrastructure (NNCI).
We additionally thank the Efimenko and Genzer groups (NCSU, CBE) for providing use of their IR-VASE and would also like to acknowledge Andrew Klump for performing ToF-SIMS measurements.

\section{Author Contributions}
E.L.R. and K.P.K. grew the ENZ-on-SPP bilayers, characterized their optical and structural properties, and simulated their optical response with the transfer matrix method. T.G.F. performed CST simulations. J-P.M. proposed the experiment and supervised the experimental work, while J.D.C. supervised the simulation work with additional analytical support from N.E. E.L.R. prepared the manuscript and figures with substantial discussion and input from all authors.

\bibliography{manuscript}
\bibliographystyle{naturemag}
\end{document}


\maketitle

\section{In:CdO bilayer growth}

Indium doped cadmium oxide (In:CdO) is grown using RF-assisted reactive
high power impulse magnetron sputtering (R-HiPIMS). Cd is sputtered
from a circular 2-inch metallic Cd target (99.9999\%, Osaka Asahi
Metal), which is pressed and cut in-house and mounted to a MeiVac
MAK 2-inch magnetron sputtering source. In is sputtered from a circular
1-inch metallic In target (99.99\%, Sigma Aldrich), which is pressed
and cut in-house and mounted to an AJA 1-inch magnetron sputtering
source. The sputtering sources are housed in a high-vacuum sputtering
system with a turbomolecular pump with a base pressure of 10\textsuperscript{-7}
Torr. All films are deposited in a combined Ar-O\textsubscript{2}
atmosphere at 10 mTorr, which is achieved by flowing 19 sccm Ar and
15 sccm O\textsubscript{2} into the vacuum chamber using mass flow
controllers and by adjusting a gate valve between the vacuum chamber
and the pump. 

Power is delivered to the Cd target using an Advanced Energy MDX 1.5K
DC power supply in concert with a Starfire Industries Impulse Pulsed
Power Module. The DC supply is run in constant voltage mode to deliver
approximately 420 V to the HiPIMS pulse generator. The HiPIMS pulsed
deposition parameters are kept constant at 80 $\mu$s pulse width
and 800 Hz repetition rate, which corresponds to a 1250 $\mu$s
pulse period and a 6.4\% sputtering duty cycle. This gives a deposition
rate of about 25 nm/minute. Doping is achieved by applying RF power
to the In target using Advanced Energy RFX-600 RF power supply with
a Manitou Systems manual matching network. Film thickness is controlled
by deposition time, while dopant/carrier concentration is controlled
by changing the RF power applied to the In target (2.4 W/cm\textsuperscript{2}
to 7.9 W/cm\textsuperscript{2}). 

All ENZ-on-SPP bilayer samples described in the main text are fabricated
by first depositing a 300 nm In:CdO layer (the SPP layer) on double-polished
epi-ready r-plane sapphire (Jiaozuo TreTrt Materials) affixed to a
stainless steel puck with silver paint (Ted Pella). The puck and substrate
are heated to a deposition temperature of 370\textdegree C as measured
using a Raytek 1.6 $\mu$m MM Series Pyrometer. In every case, this
SPP layer is grown using 6.1 W/cm\textsuperscript{2} of RF power
density on the In target to achieve carrier concentrations of 2.5$\times$10\textsuperscript{20}
cm\textsuperscript{-3}. The bilayer is completed by growing an additional
homoepitaxial layer (the ENZ layer), on top of the SPP layer, with
controlled thickness and carrier concentration. Between the SPP layer
deposition and subsequent ENZ layer deposition, the substrate is shuttered
for 5 minutes while the In target power is adjusted as desired. This
rest helps improve reproducibility in deposition rate and carrier
concentration in our bilayers. Following deposition, all samples are
annealed at 700\textdegree C in pure O\textsubscript{2} for 30 minutes. 

\section{Characterization}

Structural properties (thickness, crystallinity, etc.) of our samples
are measured by X-ray diffraction using a PANalytical Empyrean X-ray
diffractometer in parallel beam geometry (double-bounce hybrid monochromator
incident optic, parallel plate collimator receiving optic). Reciprocal
space maps are collected using a PANalytical PIXcel area detector.
In:CdO electronic properties (carrier concentration, mobility) are
measured by the Hall Effect using an Ecopia HMS-3000 system with a
0.51 T magnet. Dopant depth profiles are characterized with time-of-flight
secondary ion mass spectroscopy (ToF-SIMS) using an ION-TOF TOF-SIMS
V instrument with a cesium source for sputtering and a bismuth liquid
metal ion gun (SMIG) source in interlaces sputtering mode for analysis.
For depth profiles, a Cs\textsuperscript{+} sputtering beam with
10 keV energy and 25 nA current is rastered over a 120 by120 $\mu$m
area. The Bi\textsuperscript{3+} analysis beam is 0.4 pA at 25 keV
and rastered over a 50 by 50 $\mu$m area at the center of the sputtered
crater. The angle of incidence is 45\textdegree{}from normal for
both beams.

The plasmonic response of the bilayers is measured with infrared polarized
reflectivity spectroscopy using a Woollam IR-VASE ellipsometer. A
90\textdegree{} calcium fluoride (CaF\textsubscript{2}, Thor Laboratories)
prism, n=1.720 index matching fluid (Cargille Series M), and a home-built
aluminum sample holder are used to couple IR light into the samples
through the back side of the polished substrate in the Kretschmann-Raether
configuration. The silver paint is removed from the backside of the
substrate with a razor blade and/or sandpaper prior to optical measurements.
For each sample, the reflectivity of both p- and s-polarized light
reflectivity are collected as a function of incident light energy
and ellipsometer angle (Fig. S2a). The dispersion relationships are
plotted using $R=R_{p}/R_{s}$ as the reflectivity/color/z-coordinate,
which provides self-consistent backgrounding. Energy (in cm\textsuperscript{-1})
as is plotted as the y-coordinate, and normalized wavevector is plotted
as the x-coordinate ($k_{x}/k_{0}$ unitless, where $k_{x}$ is the
component of the incident light wavevector parallel to the film surface,
and $k_{0}=\omega/c$). To convert from ellipsometer angle to $k_{x}/k_{0}$
(Fig. S2b), a MATLAB script is used to interpolate the data over the
consistent range $1<k_{x}/k_{0}<1.3$ at each individual ellipsometer
angle and incident light energy, taking into account changes in incident
angle and momentum due to the dispersive prism and substrate. The
data are presented this way in the main text for easier visual interpretation
over the non-normalized $k_{x}$ (Fig. S2c).

\renewcommand{\thefigure}{S\arabic{figure}}
\renewcommand{\tablename}{Supplementary Table}

\begin{figure}
	\centering
	\includegraphics{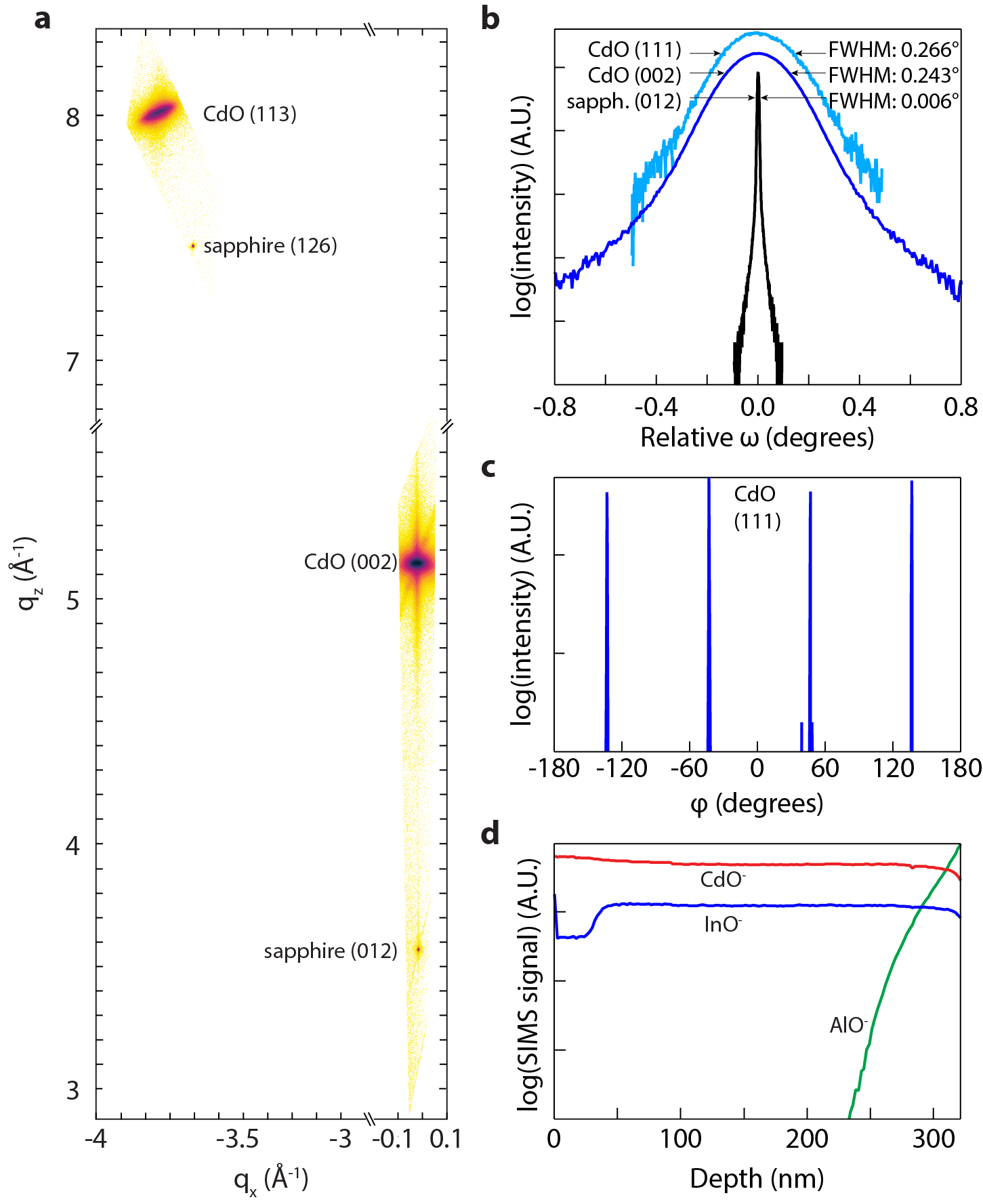}
	\caption{Structural characterization of a representative In:CdO
ENZ-on-SPP bilayer. a) Reciprocal space map showing heteroepitaxy
between CdO and the r-plane sapphire substrate. The discrete CdO (002)
and (113) peaks also indicate that the ENZ layer is homoepitaxial
with respect to the SPP layer beneath it. b) Rocking curves of the
CdO (002) and (111) reflections, and the r-plane sapphire reflection.
The low peak widths for CdO are an indication of high quality films
and hetero-/homoepitaxy. c) Skew-symmetric 360\textdegree{} $\phi$
scan about the CdO (111) reflection ($\chi$: 54.7\textdegree , $2\theta$:
33\textdegree ). The four-fold symmetry confirms hetero-/homoepitaxy.
d) ToF-SIMS spectra showing the depth profile of Cd and In in the
bilayer film (left: bilayer surface right: substrate). The InO\textsuperscript{-}
signal confirms distinct dopant concentrations in the ENZ and SPP
layers.}
  \label{fig:S1}
\end{figure}

\begin{figure}
\centering
\includegraphics{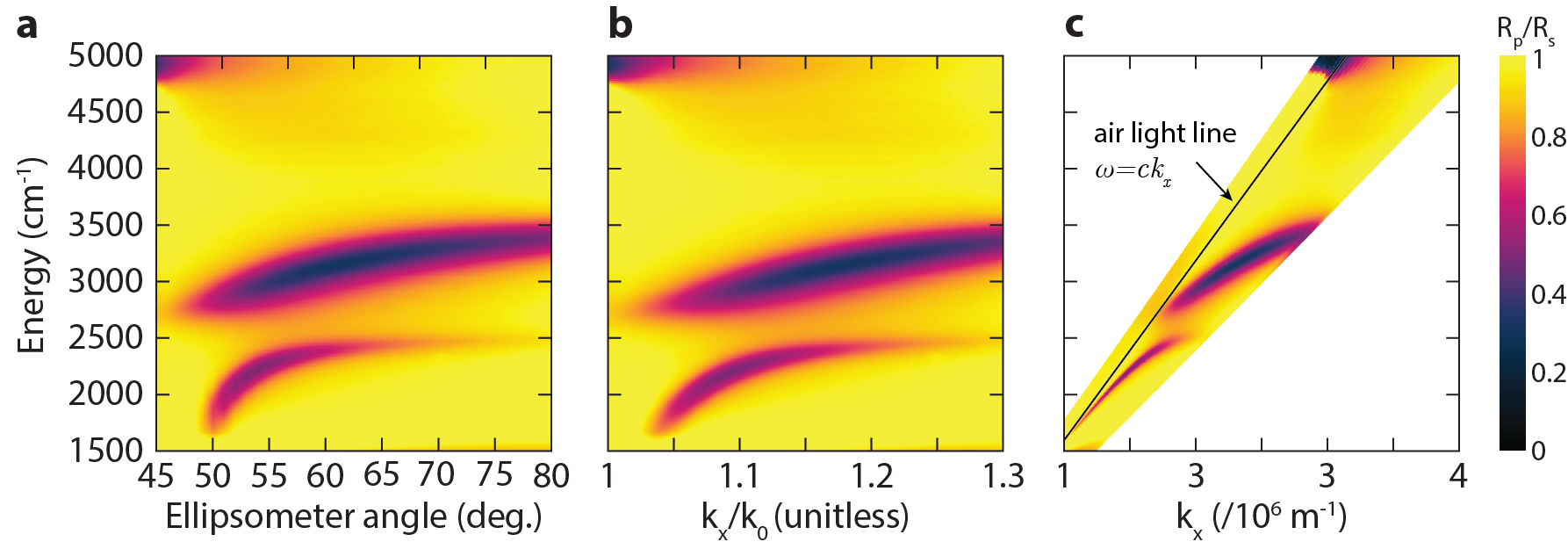}
\caption{Dispersion relationship of a PH-ENZ mode in a representative
ENZ-on-SPP bilayer plotted in terms of a) IR-VASE ellipsometer angle;
b) normalized $k_{x}/k_{0}$; and c) non-normalized $k_{x}$.}
\label{fig:S2}
\end{figure}

\section{Calculation of complex wavevector}

To extract the imaginary part of the wavevector in Figure 1g of the
main text, the reflectivity minimum as a function of wavevector is
fitted to a negative Lorentzian peak at each individual frequency
using a nonlinear curve fit algorithm in MATLAB. To achieve better
fits, a linear background component, $R_{0}-R_{1}(k_{x}/k_{0})$,
is added to the fitting function, which is:
\begin{equation}
R(k_{x}/k_{0})=R_{0}-R_{1}\left[k_{x}/k_{0}\right]-A\frac{\Gamma^{2}}{\left[k_{x}/k_{0}-k_{\mathrm{center}}\right]^{2}+\Gamma^{2}}
\end{equation}
where $R_{0}$, $R_{1}$, and $A$ are constants. The fitted Lorentzian
line width $\Gamma$ corresponds to the imaginary component of the
wavevector, while the peak location $k_{\mathrm{center}}$ gives the
real component. Given a good agreement between $k_{\mathrm{center}}$
and the experimental/simulated dispersion (Fig. S3a), propagation
lengths can then be calculated as:
\begin{equation}
\frac{1}{2\mathrm{Im}(k_{x}/k_{0})}
\end{equation}

We note that in all of these calculations the spectral dispersion
of the material is included, meaning that $k_{0}$ is calculated at
each incident frequency. Compared to the SPP reference calculations,
we see that both $\mathrm{Re}(k_{x}/k_{0})$ and $\mathrm{Im}(k_{x}/k_{0})$
diverge strongly within the middle of the gap region at the ENZ frequency
due to splitting arising from the strong modal coupling. At the edges
of the gap region, close to the ENZ frequency, both $\mathrm{Re}(k_{x}/k_{0})$
and $\mathrm{Im}(k_{x}/k_{0})$ are larger for the symmetric/antisymmetric
PH-ENZ mode than for the SPP mode due to the strong absorption and
field confinement within the ENZ layer and the associated optical
losses. 

\begin{figure}
\centering
\includegraphics{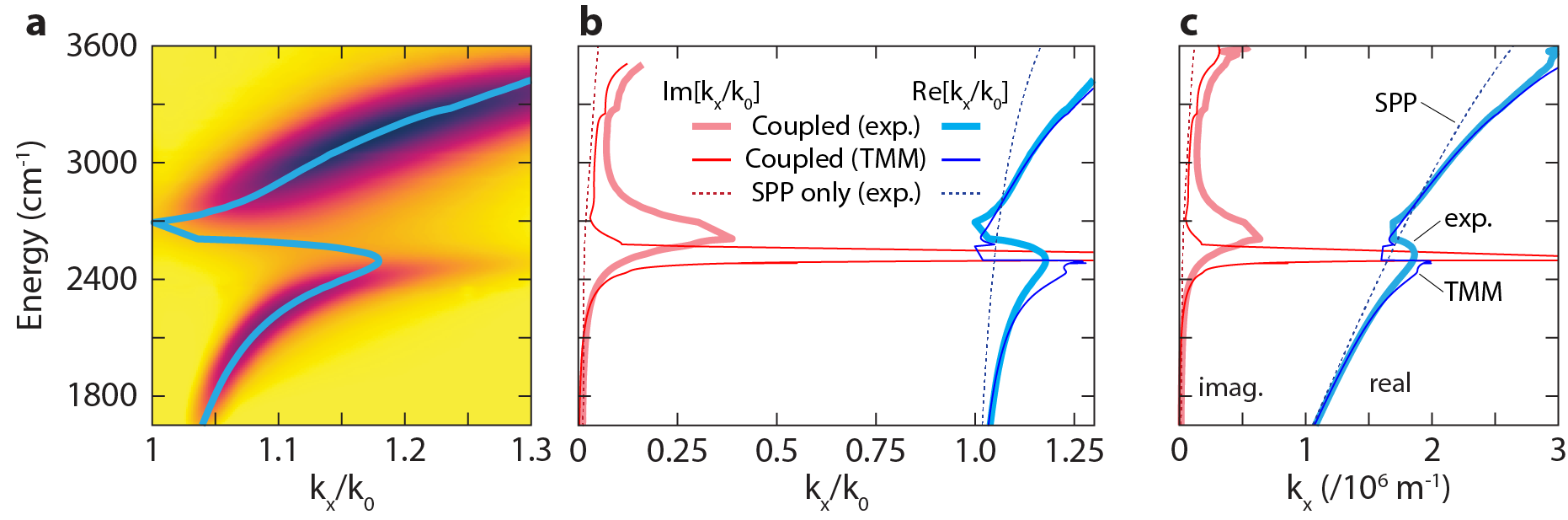}
\caption{Calculation of imaginary component of wavevector. a) Fitted
real component of wavevector ($k_{\mathrm{center}}$, blue line) overlaid
on the experimentally-measured PH-ENZ dispersion relation from Fig.
1c of the main text. b) Calculated real and imaginary wavevectors
for experimental SPP dispersion in Fig.1a, the experimental coupled
modes in Fig. 1c, and the simulated coupled modes in Fig. 1e. c)
Absolute magnitudes of the real and imaginary components of the wavevectors.}
\label{fig:S3}
\end{figure}

\section{TMM simulations}

The transfer matrix method is used to simulate the polarized reflectivity
of our samples at each discrete incident angle and energy. The transfer
matrix allows us to solve Fresnel's equations for the
multi-layered stacks examined in this study (i.e., prism//substrate//SPP
layer//ENZ layer//air) by taking advantage of the fact that the boundary
conditions must be continuous across each interface. We use the following
transfer matrices for p- and s-polarized light, respectively, moving
from layer $a$ to layer $b$:\renewcommand\arraystretch{2}

\begin{align*}
\mathbf{T}_{a\rightarrow b}^{p}= & \frac{1}{2}\left(\begin{array}{cc}
1+C_{1}C_{2} & 1-C_{1}C_{2}\\
1-C_{1}C_{2} & 1+C_{1}C_{2}
\end{array}\right)\\
\mathbf{T}_{a\rightarrow b}^{s}= & \frac{1}{2}\left(\begin{array}{cc}
1+C_{1}^{-1}C_{2} & 1-C_{1}^{-1}C_{2}\\
1-C_{1}^{-1}C_{2} & 1+C_{1}^{-1}C_{2}
\end{array}\right)\\
 & C_{1}=\sqrt{\frac{\varepsilon_{a}}{\varepsilon_{b}}};\,C_{2}=\frac{\cos\theta_{b}}{\cos\theta_{a}}
\end{align*}
where $\varepsilon_{a}(\omega)$ is the dispersive dielectric function
of layer $a$, $\varepsilon_{b}(\omega)$ is the dispersive dielectric
function of the subsequent layer $b=a+1$, and $\theta$ is the incident
light angle in layer $a$ or $b$ (corrected for the prism angle,
refraction, etc.). We use experimentally-determined dispersive dielectric
functions for the CaF\textsubscript{2} prism and sapphire substrate.
For plasmonic In:CdO layers, the dielectric function is simulated
using the Drude model: 
\begin{align*}
\varepsilon_{\mathrm{In:CdO}} & =\varepsilon_{\infty}-\frac{\omega_{p}^{2}}{\omega^{2}+i\gamma\omega}\\
\omega_{p}^{2} & =\frac{ne^{2}}{\varepsilon_{0}m_{e}^{*}}\\
\gamma & =\frac{e}{\mu m_{e}^{*}}
\end{align*}
where $\varepsilon_{\infty}$ is the high-frequency dielectric constant,
$\omega_{p}$ is the plasma frequency, $n$ is the electron concentration,
$e$ is the charge of an electron, $\varepsilon_{0}$ is the vacuum
permittivity, $m_{e}^{*}$ is the effective mass of an electron, $\gamma$
is the damping frequency, and $\mu$ is mobility.

The propagation matrix for light moving through phase $a$ is: 
\begin{align*}
\mathbf{P}_{a} & =\left(\begin{array}{cc}
\exp\left[ikd_{a}\right] & 0\\
0 & \exp\left[-ikd_{a}\right]
\end{array}\right)
\end{align*}
where $k=\frac{\omega}{c}\sqrt{\varepsilon_{a}}\cos\theta_{a}$ and
$d_{a}$ is the thickness of layer $a$. The total product matrix
for a given polarization $\mathbf{M}_{\text{pol.}}$ is the sequence
product of the transfer and propagation matrices over all layers:
\[
\mathbf{M}_{\text{pol.}}=\prod_{\text{layer }n}^{\text{all layers}}\mathbf{T}_{n-1\rightarrow n}^{\text{pol.}}\mathbf{P}_{n}\mathbf{T}_{n\rightarrow n+1}^{\text{pol.}}
\]
The reflectivity coefficient and total reflectivity for a given polarization
can be found from the product matrix by:
\begin{align*}
r_{\text{pol.}} & =\frac{m_{21}}{m_{11}}\\
R_{\text{pol.}} & =\left|r_{\text{pol.}}\right|^{2}
\end{align*}
Thus, the transfer matrix method allows us to accurately simulate
$R_{p}/R_{s}$ reflectivity maps for our bilayer samples over a discrete
range of incident angles and frequencies. The TMM simulation parameters
used in this study are tabulated below. 

\textbf{Reference}: Guske, J.T. \emph{Modeling and Prediction of Surface
Plasmon Resonance Spectroscopy.} Ph.D. dissertation (North Carolina
State University, 2013).

\begin{table}
\centering
\caption{TMM simulation parameters held constant across
all simulations}
\begin{tabular}{cc}
Parameter & Value\tabularnewline
\hline 
\hline 
$\varepsilon_{\infty}$ & 5.3\tabularnewline
$\mu$  & 300 cm\textsuperscript{2}/V·s\tabularnewline
$\gamma$  & $2.80\times10^{13}$ rad/s\tabularnewline
$m_{e}^{*}$ & $0.21\times m_{e}$\tabularnewline
SPP layer thickness & 300 nm\tabularnewline
SPP layer $n$ & $2.5\times10^{20}$ cm\textsuperscript{-3}\tabularnewline
SPP layer $\omega_{p}$  & $1.95\times10^{15}$ rad/s\tabularnewline
\end{tabular}

\end{table}

\begin{table}
\centering
\caption{TMM simulation parameters varied across simulations,
organized by appearance in the main text.}
\begin{tabular}{cccc}
Fig. reference & ENZ layer thickness & ENZ layer $n$ (cm\textsuperscript{-3}) & ENZ layer $\omega_{p}$ (rad/s)\tabularnewline
\hline 
\hline 
1e & 100 nm & $8.1\times10^{19}$ & $1.11\times10^{15}$\tabularnewline
2a & 100 nm & $3.5\times10^{20}$ & $2.30\times10^{15}$\tabularnewline
2b & 100 nm & $1.5\times10^{20}$ & $1.51\times10^{15}$\tabularnewline
2c & 100 nm & $9.5\times10^{19}$ & $1.20\times10^{15}$\tabularnewline
2d & 100 nm & $4.0\times10^{19}$ & $7.79\times10^{14}$\tabularnewline
3b & 80 nm & $1.5\times10^{20}$ & $1.51\times10^{15}$\tabularnewline
3e & 20 nm & $1.5\times10^{20}$ & $1.51\times10^{15}$\tabularnewline
\multirow{2}{*}{5} & \multirow{2}{*}{5-300 nm} & \multirow{2}{*}{$3-25\times10^{19}$ (increment: $1\times10^{19}$)} & \multirow{2}{*}{$6.74\times10^{14}-1.95\times10^{15}$}\tabularnewline
 &  &  & \tabularnewline
\end{tabular}

\end{table}

\section{CST simulations}

Full numerical simulations of the electromagnetic field are conducted
using the frequency domain solver CST Studio Suite 2017. Plane waves
were used to excite a 4$\mu$m by 4$\mu$m unit cell, consisting
of a calcium fluoride background, a sapphire substrate (1$\mu$m),
doped CdO layers (thicknesses same as in experiment), and then a vacuum
spacer (7.5$\mu$m). The compressed substrate layers ensure accurate
in-plane wavevectors without excessive computational cost. Reflectivity
maps are produced by varying incident angle ($\theta$), and then
converting to wavevector as described above. Two-dimensional field
profiles were extracted from the center of the simulation at the particular
frequencies and angles described in the main text.

\begin{figure}
\centering
\includegraphics[width=1\textwidth]{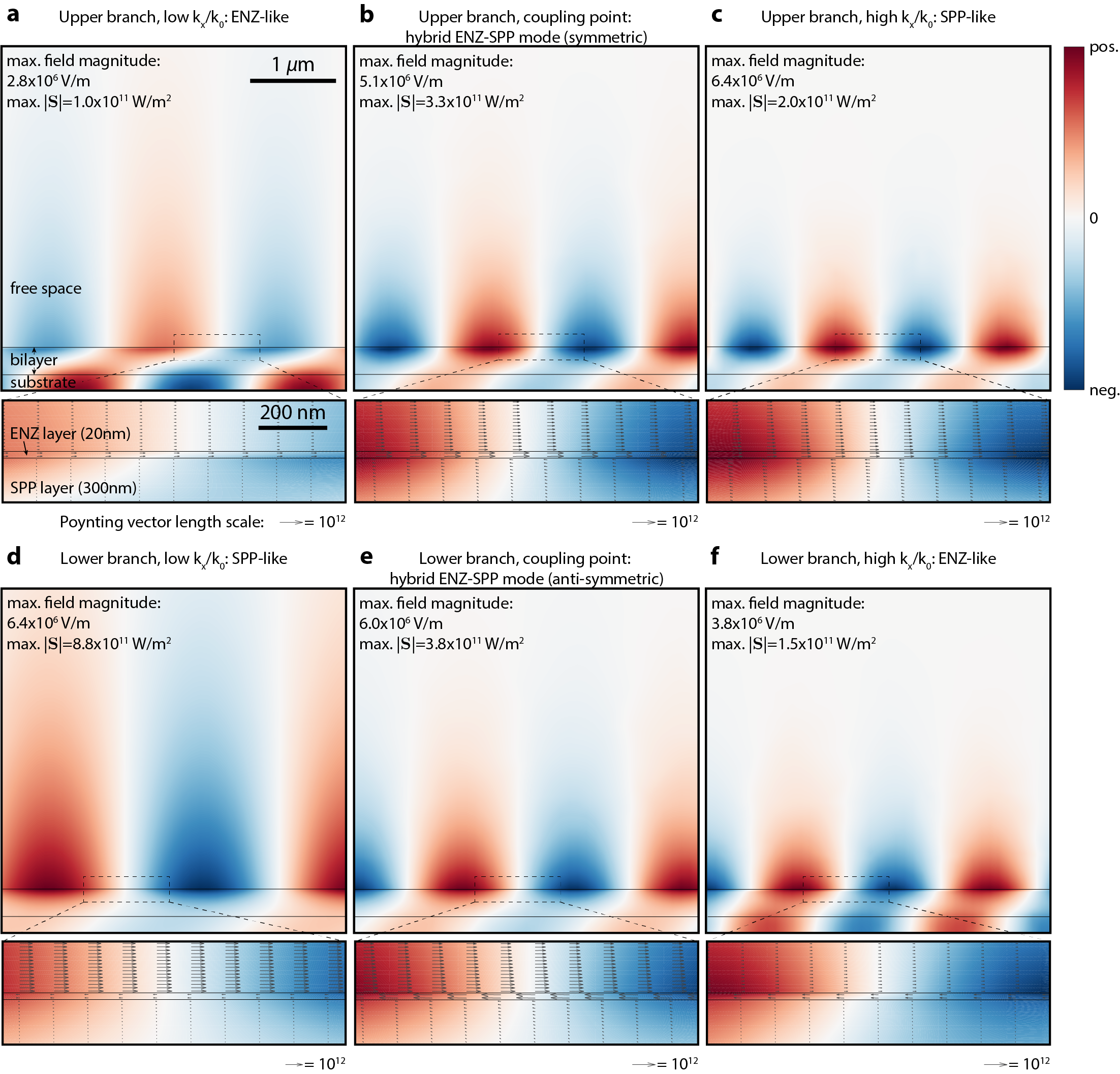}
\caption{Electric field profiles of the ENZ-on-SPP bilayer simulated
in Fig. 3f (see structure in Fig. 3d). The color axis corresponds
to the electric field in the longitudinal/x-direction, and the Poynting
vector field is overlaid in each detail view. a) Uncoupled upper branch
mode profile (point a in Fig. 3f). b) Strongly coupled upper branch
PH-ENZ mode profile (point b in Fig. 3f). c) Uncoupled upper branch
mode profile (point c in Fig. 3f) with predominant SPP character.
Detail view: Evanescent field decay into free space and into the bulk
of the SPP layer with lowered propagation. d) Uncoupled lower branch
mode profile (point d in Fig. 3f) with predominant SPP character.
e) Strongly coupled upper branch PH-ENZ mode profile (point e in Fig.
3f) with hybrid ENZ-SPP behavior. f) Uncoupled lower branch mode profile
(point f in Fig. 3f) with predominant ENZ character.}
\label{fig:S4}
\end{figure}